\documentclass[aps,prl,preprintnumbers,nofootinbib,superscriptaddress,fleqn,floatfix,tightenlines,twocolumn,10pt]{revtex4-2} 
\usepackage{amsmath,amsfonts,amssymb,amscd,amsxtra,amsthm}
\usepackage{graphicx}  
\usepackage{epstopdf}
\usepackage{dcolumn}  
\usepackage{bm}          
\usepackage{slashed}
\usepackage{cancel} 
 \usepackage{float}
\usepackage{mathtools}
\usepackage{amsbsy}
\usepackage{amstext}

\usepackage[utf8]{inputenc} 
\usepackage{booktabs} 
\usepackage[normalem]{ulem} 
\usepackage[dvipsnames]{xcolor} 
\usepackage{tabularx}
\usepackage{enumitem}  
\usepackage{array} 
\usepackage{slashed}
\usepackage{tikz}
\usepackage{multirow}
\usepackage{cancel}
\usepackage{physics}

\renewcommand{\sout}{\bgroup \color{red} \ULdepth=-.5ex \ULset}

\makeatletter

\begin{document}
\preprint{INHA-NTG-04/2023}
\title{Role of strange quarks in the $D$-term and cosmological
  constant term of the proton}   
\author{Ho-Yeon Won}
\email[E-mail: ]{hoywon@inha.edu}
\affiliation{Department of Physics, Inha University, Incheon 22212,
South Korea}
 
\author{Hyun-Chul Kim}
\email[E-mail: ]{hchkim@inha.ac.kr}
\affiliation{Department of Physics, Inha University, Incheon 22212,
South Korea}
\affiliation{School of Physics, Korea Institute for Advanced Study
(KIAS), Seoul 02455, South Korea}

\author{June-Young Kim}
\email[E-mail: ]{jykim@jlab.org}
\affiliation{Theory Center, Jefferson Lab, Newport News, VA 23606, USA} 
\date {\today}
\begin{abstract}
We investigate the mechanics of the proton by examining the
flavor-decomposed proton cosmological constants and generalized vector
form factors. The interplay of up, down, and strange quarks within the
proton is explored, shedding light on its internal structure. The
contributions of strange quarks play a crucial role in the $D$-term
and cosmological constants. We find that the flavor blindness of the
isovector $D$-term form factor is only valid in flavor SU(3) symmetry.
\end{abstract}
\maketitle

\textit{Introduction} --
The proton, a fundamental building block of matter, possesses a set of
fundamental observables, including its electric charge, magnetic
dipole moment, mass, spin, and the $D$-term. The $D$-term, akin to
these observables, plays a crucial role in unraveling the mechanical
properties of the proton and shedding light on how it achieves
stability through the intricate interplay of quarks and
gluons. Understanding the distribution of mass, spin, pressure, and
shear force within the proton is facilitated by its gravitational form
factors (GFFs)~\cite{Polyakov:2002yz} in a manner similar to how
the electromagnetic form 
factors reveal charge and magnetic distributions. Specifically, the
pressure and shear-force distributions are intimately linked to the
$D$-term form factor (see a recent review and references
therein~\cite{Polyakov:2018zvc}). 

Direct measurement of the proton GFFs necessitates the interaction of
gravitons with protons, which is experimentally impractical due to the
exceedingly weak gravitational coupling strength of the
proton. However, a promising avenue emerges through the generalized
parton distributions (GPDs), which offer indirect access to the
mechanical properties of the proton. The GFFs can be regarded as the
second Mellin moments of the vector GPDs ~\cite{Muller:1994ses,
  Ji:1996nm, Radyushkin:1996nd} (see also reviews~\cite{Diehl:2003ny,
  Belitsky:2005qn}), providing insight into the proton's mechanical
structure~\cite{Polyakov:2018zvc}. Deeply virtual Compton scattering 
(DVCS) serves as an effective means to access the GPDs, enabling the
extraction of valuable information about the
GFFs~\cite{Kumericki:2009uq, Muller:2013jur, Kumericki:2016ehc}. 

Recent advancements by Burkert et al. ~\cite{Burkert:2018bqq} have
witnessed the experimental extraction of the quark component of the
proton $D$-term form factor, marking a significant breakthrough. By
leveraging experimental data on the beam-spin asymmetry and
unpolarized cross-section for DVCS on the proton, they successfully
obtained valuable insights into the $D$-term. However, their analysis
assumed the large $N_c$ limit, thereby considering the up-quark
contribution ($d_1^u$) to be approximately equal to the down-quark
contribution ($d_1^d$) in the leading order. This assumption leads to
an almost null result for the leading isovector $D$-term
$d_1^{u-d}$~\cite{Burkert:2023wzr}. Hence, it becomes imperative to
critically examine the validity of this assumption. DVCS provides an
effective way to access the GPDs. 

Furthermore, Burkert et al. ~\cite{Burkert:2018bqq} also assumed
flavor SU(2) symmetry, neglecting the contribution of strange
quarks. However, as we shall establish in this study, the inclusion of
strange quarks becomes indispensable for accurately describing the
$D$-term. While the strange quark's contribution to the nucleon mass
and spin may be marginal, it assumes a pivotal role in characterizing
the $D$-term, indicating that the nucleon's stability can only be
comprehensively understood by considering the degrees of freedom
associated with up, down, and strange quarks. Note that the gluon
GPDs are only accessible at higher orders in $\alpha_s$, so they are 
expected to be smaller than the quark GPDs.

In this Letter, our objective is to elucidate the significance of
strange quarks in unraveling the mechanical structure of the
proton within the framework of the chiral quark-soliton model
($\chi$QSM) ~\cite{Diakonov:1987ty, Christov:1995vm,Diakonov:1997sj},
The $\chi$QSM provides a suitable relativistic quantum-field
theoretic framework for our analysis. It is noteworthy that previous
studies employing the $\chi$QSM~\cite{Goeke:2007fp} have yielded
results in excellent agreement with experimental data on the pressure
and shear-force distributions of the proton, as reported by Burkert et 
al.~\cite{Burkert:2018bqq}.  In a recent publication by Won et
al. ~\cite{Won:2023rec}, the proton $D$-term and PCC were further
explored by decomposing them into up and down-quark flavor components
through the computation of generalized isovector vector form
factors. The magnitude of the down-quark component was found to
be larger than that of the up-quark component, leading to a nonzero
value of $d_1^{u-d}$ (see also Ref.~\cite{Wakamatsu:2007uc}). However,
in order to comprehensively understand the proton's mechanical
structure, it becomes imperative to consider flavor SU(3) symmetry and
incorporate the generalized triplet and octet vector form factors
alongside the GFFs. This extension enables us to decompose the GFFs
into their up, down, and strange quark components, thereby gaining a
deeper understanding of the internal mechanics of the proton.  
 
Notably, a recent lattice calculation has provided insights into the
flavor decomposition of the proton's spin and 
momentum~\cite{Alexandrou:2017oeh}. However, it is important to
emphasize that the flavor-decomposed PCCs were not 
considered in this analysis. In our current work, we aim to fill this
gap by investigating the role of flavor-decomposed PCCs in examining
the mechanical structure of the proton. By incorporating these crucial
factors, we can refine our understanding of the proton's intricate
mechanics and shed further light on its fundamental properties. 

\vspace{0.5cm}
\textit{General vector form factors} -- 
The GFFs can be related to the matrix element of the
flavored $(q)$ symmetric energy-momentum tensor~(EMT) current defined as  
$\hat{T}_q^{\mu\nu}=\bar{q}\frac{i}{4}
\overleftrightarrow{D}^{\{\mu} \gamma^{\nu\}}q$    
with the covariant derivative $\overleftrightarrow{D}^{\mu} =
\overleftrightarrow{\partial}^{\mu}-2igA^{\mu}$
and $\overleftrightarrow{\partial}^{\mu} =
\overrightarrow{\partial}^{\mu}-\overleftarrow{\partial}^{\mu}$ , and
$a^{ \{\mu }b^{  \nu \}} = a^{\mu} b^{\nu} + a^{\nu} b^{\mu}$, which 
can be parametrized in terms of the four GFFs $A_q$, $J_q$, 
$D_q$, and $\bar{c}_q$:
\begin{align}
&   \mel{p'}{   \hat{T}_q^{\mu\nu}  ( 0 )  }{p}   \cr 
&   \hspace{-0.7cm} = \bar{u}(p')
    \Bigg[  A_q  ( t )  \frac{  P^{\mu} P^{\nu} }{  M_{N} } 
   + J_q  ( t ) \frac{  i ( P^{\mu} \sigma^{\nu\rho} + P^{\nu}
    \sigma^{\mu\rho}  )   \Delta_{\rho} } { 2 M_{N} }   \cr
&   \hspace{-0.7cm}
  + D_q  ( t )     \frac{  \Delta^{\mu} \Delta^{\nu} 
  - g^{\mu\nu} \Delta^{2}}{4M_{N}}   + \bar{c}_q ( t ) M_{N}
    g^{\mu\nu} \Bigg]  u(p), 
\label{eq:1}
\end{align}
where $P= (p'+p)/2$, $\Delta= p'-p$, and $\Delta^{2} = -\bm{\Delta}^{2}= t$.
$A_q$, $J_q$, and $D_q$ are related to the second
moments of the vector GPDs defined in Ref.~\cite{Ji:1996nm}
\begin{align}
 A_q(t) &= A_{20,q}(t), \;
2J_q(t) =  A_{20,q}(t)+B_{20,q}(t), \cr
               D_q(t) &=  4C_{20,q}(t),
\label{eq:2_1}                             
\end{align}
where the subscript stand for the quark flavor. In the forward limit
($t\to 0$)~\cite{Lorce:2017xzd, Lorce:2021xku}, Eq.~\eqref{eq:1}
reduces to  
\begin{align}
  \label{eq:2}
\hspace{-0.8cm} \mel{p}{   \hat{T}_q^{\mu\nu} }{p} 
 = \bar{u}(p)
\Bigg[  A_q  ( 0 )  \frac{  p^{\mu} p^{\nu} }{  M_{N} }
+ \bar{c}_q ( 0 ) M_{N}
    g^{\mu\nu} \Bigg]  u(p)   .                                                           
\end{align}

In the $\chi$QSM, the gluon degrees of freedom were integrated out
through the instanton vacuum, and their effects are absorbed in the
dynamical quark mass $M$, which was originally
momentum-dependent~\cite{Diakonov:1985eg, Diakonov:2002fq}.
Thus, the quark part of the EMT current is conserved within the
framework of the $\chi$QSM
\begin{align}
\sum_{q}  \partial_{\mu} \hat{T}_q^{\mu\nu}  = 0.
\label{eq:3}
\end{align}
It implies that the PCC form factor $\bar{c}=\sum_q \bar{c}_q$
vanishes in the whole range of the momentum transfer. At $t=0$, the
mass, spin, and cosmological constant of the proton are normalized 
respectively as $A=1$, $J=\frac{1}{2}$, and $\bar{c}=0$. Note that,
however, there are no constraints on the generalized triplet and octet
vector form factors. 

\vspace{0.5cm}
\textit{The chiral quark-soliton model: pion mean-field approach} --
The $\chi$QSM has been 
successful in describing not only the well-known baryonic 
observables~\cite{Christov:1995vm, Kim:2018cxv} 
but also the first data on the $D$-term form
factor~\cite{Burkert:2018bqq, Burkert:2023wzr} and other
GFFs~\cite{Goeke:2007fp, Kim:2020nug, Won:2022cyy}.  
The formalism is well known already~\cite{Won:2022cyy}, we briefly
explain the essential feature of the model. The detailed expressions
can be found in a forthcoming work~\cite{Won:2023prep}.
We start from the low-energy QCD effective partition function in
Euclidean space~\cite{Diakonov:1985eg, Diakonov:1987ty,
  Christov:1995vm, Diakonov:2002fq}   
\begin{align}
  \hspace{-0.6cm}  \mathcal{Z}_{\mathrm{eff}}
&= \int  \mathcal{D} \pi^a \exp\left[  - S_{\mathrm{eff}}
  ( \pi^a )  \right], 
\end{align}
where $\pi^a$ is the SU(3) pseudo-Nambu-Goldstone (pNG) boson fields
with the superscript $a=1,\cdots 8$ and $S_{\mathrm{eff}}$ represents
the effective chiral action expressed as   
\begin{align}
S_{\mathrm{eff}} = -N_c \mathrm{Tr}\log\left[
i\slashed{\partial} + i M U^{\gamma_5} + i\hat{m}\right]. 
\end{align}
$N_c$ denotes the number of colors. The chiral field $U^{\gamma_5}$ is
defined by $U^{\gamma_5} :=e^{i\gamma_5 \pi^a \lambda^a}= P_LU + P_R
U^\dagger$, where $P_{L(R)}:=(1\mp\gamma_5)/2$ and $U:=e^{i \pi^a
  \lambda^a}$. $M$ stands for the dynamical quark mass, which arises
from the spontaneous breakdown of chiral
symmetry~\cite{Diakonov:1985eg,Diakonov:2002fq}. Though $M$ is
originally momentum-dependent and its value at the zero virtuality is
determined by the saddle-point equation from the instanton 
vacuum~\cite{Diakonov:1985eg,Diakonov:2002fq}, we will use it as the
only free parameter in the current work. However, $M=420$ MeV is known 
to be the best value for describing various baryonic
observables~\cite{Christov:1995vm, Kim:2018cxv}. 
$\hat{m}$ is the current-quark mass matrix
$\mathrm{diag}(m_{\mathrm{u}} \,m_{\mathrm{d}},\,m_{\mathrm{s}})$. We
consider and flavor SU(3) symmetry ($m_{\mathrm{u}} =
m_{\mathrm{d}}=m_{\mathrm{s}}$)  in the current work.   

The first three components of the pNG fields can be coupled to the
three dimensional coordinates, which is called the hedgehog ansatz
$\pi^a = P(r) n^a$ with the unit basis vectors $n^a=x^a/r$. It is the
minimal generalization that allows to incorporate the pion fields.
$P(r)$ is called the profile function for the classical soliton. It
can be determined by solving the classical equation of motion 
self-consistently. In flavor SU(3) symmetry, we employ the Witten's
embedding to preserve the hedgehog symmetry
\begin{align}
U = e^{i \pi^a \lambda^a} = \begin{pmatrix}
e^{i\bm{n}\cdot \bm{\tau} P(r)}  & 0 \\
  0 & 1 \\
\end{pmatrix}.
\end{align}
Note that the zero-mode quantization with this embedding correctly
yields the spectrum of the lowest-lying SU(3) baryons such as the
baryon octet and decuplet. The zero-mode quantization can be performed
by the functional integration over rotational and translational zero
modes of the $U$ field. Including the external tensor
source field, we can evaluate the matrix element of the EMT current.  
The zero-mode quantization naturally furnishes the rotational $1/N_c$
corrections. While they do not contribute to the GFFs except for $J_q(t)$,
they provide substantial effects on the generalized triplet and octet
vector form factors. 

We present the final expressions for the GFFs with flavor $q$:
($\chi=3,\,8$) 
\begin{align}
&   \hspace{-0.3cm}\left[
    A_q(t) + \bar{c}_q (t)
  - \frac{t}{4M_{N}^{2}}\left(D_q  ( t )  - 2 J_q (  t ) \right) 
    \right] \cr
  & = \frac{4\pi}{M_{N}} \int dr\, r^2
    j_{0}(kr)\varepsilon_q \left(r\right)  ,\cr
&  \hspace{-0.3cm} \left[    \bar{c}_q (t)  - \frac{t}{6M_{N}^{2}} 
    D_{q}  \right] = - \frac{4\pi}{M_{N}} \int   dr\, r^2
  j_{0}(kr) p_q \left(r\right)  ,\cr
      &   D_q (t) =
        16\pi M_{N} \int dr \,r^2 \frac{j_{2}(kr)}{t} s_q
        \left(r\right)  ,\cr
&    J_q (t)
  = 12 \pi\int dr\,r^2 \frac{j_{1}(kr)}{kr}
    \rho_q^{J} \left(r\right)  ,
    \label{eq:8}
\end{align}
with $k=\sqrt{-t}$ and $J_3'=J_3=1/2$. For detailed expressions for the
distributions, we refer to the forthcoming work~\cite{Won:2023prep}.

\begin{figure}[htp]
  \centering
  \includegraphics[scale=0.33]{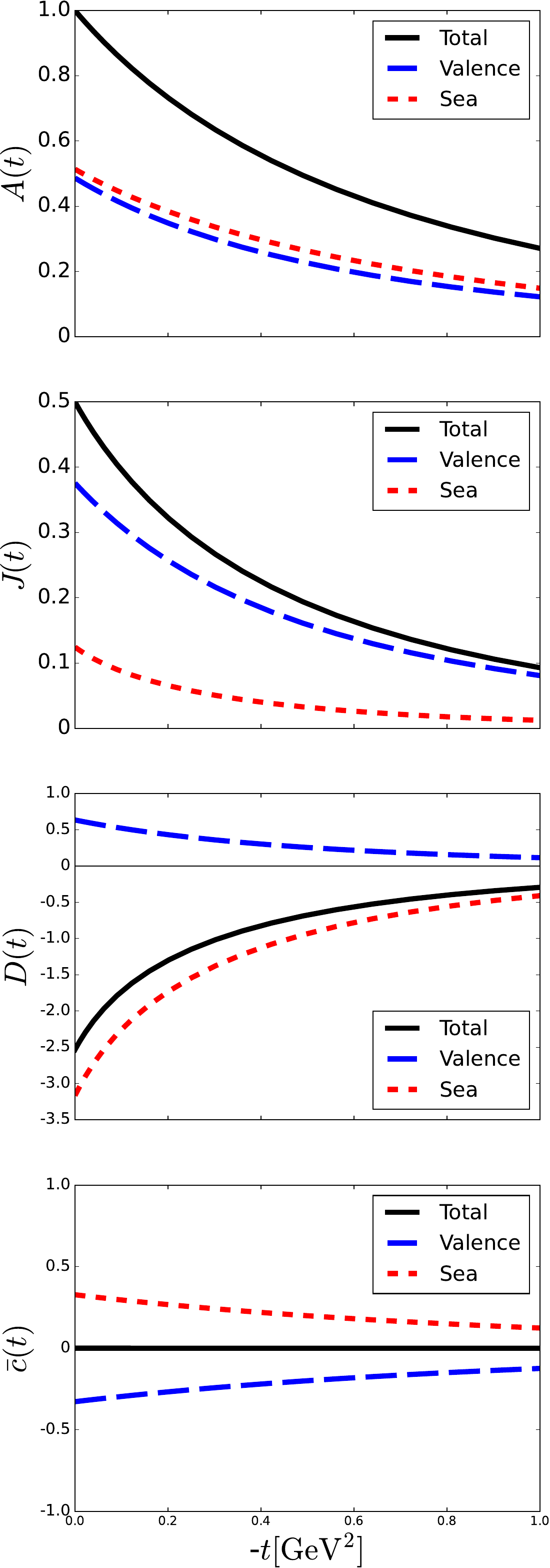}
  \caption{Results for the gravitational form factors of the proton.
 The dashed, short-dashed, and solid curves draw the valence-quark,
 sea-quark, and total contributions, respectively.}    
 \label{fig:1}
\end{figure}

\vspace{0.5cm}
\textit{Results and discussion} --
Figure~\ref{fig:1} illustrates the results obtained for the GFFs. The
dashed curves represent the contributions from valence quarks, while
the short-dashed curves depict the contributions from sea
quarks. Remarkably, the sea-quark contributions are found to dominate
over the valence-quark contributions, as demonstrated in the first
panel of Fig.~\ref{fig:1}. This observation 
aligns with the classical nucleon mass values, given by
$M_N = N_{c} E_{\mathrm{val}} + E_{\mathrm{sea}} =
611.1\,\mathrm{MeV} + 645.3\,\mathrm{MeV} = 1256$ MeV. It is
noteworthy that sea quarks contribute approximately $51.4~\%$ to the
proton's mass.  

Regarding the average momentum fraction, a recent lattice calculation
yielded a value of $\langle x\rangle_p = 0.497(12)(5)|_{\mathrm{conn}}
+ 0.307(121)(95)|_{\mathrm{disc}} + 0.267(12)(10)|_{\mathrm{gluon}} =
1.07(12)(10)$~\cite{Alexandrou:2017oeh}, which can be identified as 
the mass form factor $A(t)$ in the forward limit. By assuming that the 
sea-quark contributions implicitly incorporate the effects of
integrated-out gluon degrees of freedom within the instanton vacuum,
our current findings are in good agreement with the lattice data. 
In contrast to the proton mass, the sea quarks contribute
approximately $24~\%$ to the proton spin, as depicted in the second
panel of Fig.~\ref{fig:1}.

In the third panel of Fig.~\ref{fig:1}, it is evident that the
sea-quark contribution surpasses the valence-quark contribution. This
observation aligns with the nature of the proton $D$-term form
factor, which exhibits a quadrupole structure due to the rank-2 tensor
characteristics of the EMT, as described in
Eq.~\eqref{eq:8}. Remarkably, similar behavior can be observed in the
electric quadrupole (E2) form factors of spin-3/2 baryons, as
discussed in Refs.~\cite{Kim:2019gka, Kim:2020lgp, Kim:2020uqo,
  Kim:2021xpp}.

The PCC form factor $\bar{c}(t)$ is
expected to vanish, as a consequence of EMT current conservation, as
shown in Eq.~\eqref{eq:3}. Interestingly, the valence-quark
contribution is exactly canceled by the sea-quark contribution. These
results, depicted in Fig.~\ref{fig:1}, underscore the importance of
employing relativistic quantum-field theoretic approaches to
comprehend the mechanical structure of the proton. Such methodologies
are crucial for unraveling the intricate dynamics within the proton
and gaining deeper insights into its mechanical properties. 

\begin{figure}[htp]
  \centering
  \includegraphics[scale=0.33]{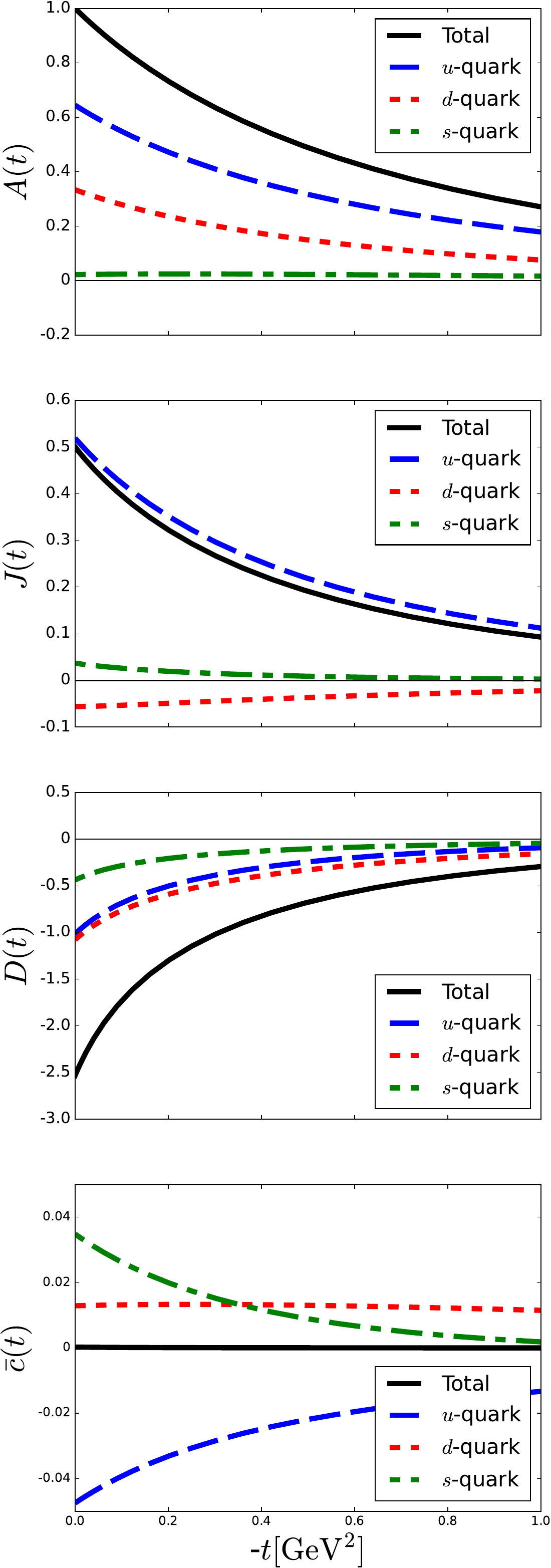}
  \caption{Results for the gravitational form factors of the proton.
  The long-dashed, short-dashed, dot-dashed, and solid curves  draw the
  up-quark, down-quark, strange-quark, and total contributions,
  respectively. 
}  
\label{fig:2}
\end{figure}
Once we have evaluated the GFFs and we can decompose 
the GFFs as follows: 
\begin{align}
 F_{u} &=F^0/3+F^3/2 + F^8/2\sqrt{3}, \cr
F_{d} &=F^0/3-F^3/2 + F^8/2\sqrt{3}, \cr
 F_{s} &=F^0/3 - F^8/\sqrt{3},
   \label{eq:9}
\end{align}
where $F^0$, $F^3$, and $F^8$ denote the generic GFFs, and generalized
triplet and octet form factors, respectively.
The flavor decomposition of the proton mass form factor $A(t)$ is
presented in the upper panel of Fig.~\ref{fig:2}. As emphasized in the
Introduction, considering the PCCs is crucial for a proper
understanding of the proton mass decomposition~\cite{Lorce:2017xzd,
  Lorce:2021xku}. The contribution of strange quarks to the mass form
factor is found to be negligible. Again, the up-quark contribution dominates
over the contributions from down and strange quarks. This dominance of
up quarks is also reflected in the  proton spin, with up quarks
accounting for the majority inside a proton, as depicted in the
second panel of Fig.~\ref{fig:2}. In contrast, the spin of the neutron
is primarily attributed to down quarks, as evidenced in Table~\ref{tab:1}. 

\begin{table*}[htp] 
  \caption{Results for the flavor-decomposed GFFs of the proton at $t=0$}
\begin{center}
  \renewcommand{\arraystretch}{1.7}
\scalebox{1.1}{%
\begin{tabular}{ccccccccccccc} 
  \hline
  \hline
  $N$    
  & $A_{u}  ( 0 ) $  & $A_{d} ( 0 ) $ & $A_{s} ( 0 ) $  
  & $J_{u}  ( 0 ) $  & $J_{d} ( 0 ) $ & $J_{s} ( 0 ) $  
  & $D_{u} ( 0 ) $  & $D_{d} ( 0 ) $ & $D_{s} ( 0 ) $         
  & $\bar{c}_{u} ( 0 ) $  & $\bar{c}_{d} ( 0 ) $ & $\bar{c}_{s} ( 0 ) $         \\
  \hline
  $p$          
  &  $0.644$ &  $0.334$ &  $0.022$ 
  &  $0.519$ & $-0.056$ &  $0.037$ 
  & $-1.016$ & $-1.077$ & $-0.438$ 
  & $-0.048$ &  $0.013$ &  $0.035$ \\
  $n$          
  &  $0.334$ &  $0.644$ &  $0.022$ 
  & $-0.056$ &  $0.519$ &  $0.037$ 
  & $-1.077$ & $-1.016$ & $-0.438$ 
  &  $0.013$ & $-0.048$ &  $0.035$ \\
  \hline
  \hline
\end{tabular}}
\end{center}
\label{tab:1}
\end{table*}
The flavor decomposition results for the $D$-term form factor are
displayed in the third panel of Fig.~\ref{fig:2}. Notably, the up- and
down-quark contributions exhibit remarkable similarity, while the
strange-quark contribution comprises approximately $25~\%$ of their 
combined effect. This finding exhibits profound physical implications.  
The result for the flavor-decomposed $D$-term shown in the third panel
of Fig.~\ref{fig:2} apparently yields almost a neglible value of $D^{u-d}$.  
However, one should keep in mind that in flavor SU(2) $D^{u-d}$ is
nonnegligible~\cite{Won:2023rec}. It implies that a certain amount of
the $d$-quark contribution is taken over by the strange quark in
flavor SU(3). Thus, the blindness of $D^{u-d} \sim 0$ assumed in
Ref.~\cite{Burkert:2018bqq} is only valid in the flavor SU(3) symetric
case.  

In the final panel of Fig.~\ref{fig:2}, the flavor decomposition of the
PCC form factor is depicted. Interestingly, the up and strange-quark
contributions dominate the PCC form factor, while the down-quark
contribution is relatively small. The PCC form factor $\bar{c}^0$
vanishes due to the conservation of the EMT current, whereas
$\bar{c}^3$ and $\bar{c}^8$ can have finite values. Notably, both
$\bar{c}^3$ and $\bar{c}^8$ exhibit negative numerical values,
resulting in a small magnitude for $\bar{c}_{d}$ according to
Eq.~\eqref{eq:9}. On the other hand, the up and strange components are
defined respectively as $(\bar{c}^3+\bar{c}^8/\sqrt{3})/2$ and
$-\bar{c}^8/\sqrt{3}$. Although the sign of $\bar{c}_{s}$ is opposite
to $\bar{c}_{u}$, its magnitude is comparable to that of the up-quark
component. This observation has significant implications. While the
PCC form factor itself vanishes, its flavor-decomposed components
remain finite and the strange quark comes into a crucial role. As
stated in Eq.~\eqref{eq:8}, the PCC is linked to the 
pressure distribution, indicating that the strange quark should be
considered in understanding the internal mechanical structure of the
proton. Furthermore, it suggests that when the $D$-term form factor is
extracted in future experimental data, one should carefully consider the
contribution from strange quarks. 

When the GFFs are understood as the second Mellin moments of the
vector GPDs as expressed in Eq.~\eqref{eq:2_1}, the PCC does not
appear from the leading-twist GPDs. The form factor $A_q$ in the
forward limit is identified as the momentum fraction of the proton,
denoted as $\langle x\rangle_q$. However, if we specifically consider
the temporal component of Eq.~\eqref{eq:2}, we derive the general
decomposition of the proton mass in the rest frame
as~\cite{Lorce:2017xzd, Lorce:2021xku}   
\begin{align}
  \label{eq:10}
M_p = \sum_q (A_q(0)+ \bar{c}_q(0)) M_p,
\end{align}
which leads to $\sum_q (A_q(0)+ \bar{c}_q(0))=1$. 

Given the conservation of the EMT current, the total PCC is expected
to vanish, implying that $\sum_q A_q = 1$. However, it does not
necessarily mean that each flavor component $\bar{c}_q$ is
zero. Consequently, the decomposed momentum fraction, denoted as 
$\langle x\rangle_q$, may not be equivalent to the decomposed proton
mass expressed as $M_p^q=(A_q(0) +\bar{c}_q(0))M_p$. Hence, a
compelling comparison arises between $M_p^q$ and $\langle x \rangle_q$:  
\begin{align}
  \label{eq:6}
M_p^{\mathrm{u}}/M_p &= 59.7\,\% < \langle x \rangle_{u} =
                   64.4~\%,   \cr
M_p^{\mathrm{d}}/M_p &= 34.6\,\% > \langle x \rangle_{d} =
                   33.4~\%,   \cr
M_p^{\mathrm{s}}/M_p &= 5.7\,\% > \langle x \rangle_{s} =
                   2.2~\%.
\end{align}
These results indicate a remarkable feature of $\bar{c}_q$ in
describing the proton mass. One can generalize the above findings as
follows: if the $\bar{c}_{q}$ is positive (negative), 
then the flavor-decomposed proton $M^{q}_{p}/M_{p}$ is larger
(smaller) than the nucleon momentum fraction carried by quarks
$\langle x \rangle_{q}$:  
\begin{align}
    \bar{c}_{q}(0)>0 \quad \to \quad M^{q}_{p}/M_{p} > \langle x
  \rangle_{q}, \cr 
    \bar{c}_{q}(0)<0 \quad \to  \quad M^{q}_{p}/M_{p} < \langle x
  \rangle_{q}. 
\end{align}
If the $\bar{c}_{q}(0)$ is zero, then we obtain the trivial relation, 
i.e., $M^{q}_{p}/M_{p} = \langle x \rangle_{q}$.

\vspace{0.5cm}
\textit{Summary and conclusions} --
We have conducted an investigation into the flavor decomposition of
the gravitational form factors of the proton. Here are the key
findings of our study:
\begin{itemize}
\item The dominant effects on the mass, $D$-term, and cosmological
  constant of the proton stem from the sea quarks rather than the
  valence quarks. This emphasizes the necessity of employing a
  relativistically quantum-field theoretic approach to accurately
  describe these quantities.
\item The contributions of strange quarks play a particularly
  significant role in the $D$-term and cosmological constants.
 Therefore, when extracting these contributions from
  experimental data, it is essential to take into account the
  influence of strange quarks.
\item While the mass form factor $A(t)$ is associated with the average
  momentum fraction, the decomposition of the nucleon mass reveals a
  noteworthy contribution from the cosmological constants. This
  indicates a profound connection between the mass distribution of the
  proton and its mechanical structure.   
\item Burkert et al.~\cite{Burkert:2018bqq} assumed the flavor
  blindness of the isovector $D$-term ($D^{u-d}\sim 0$). 
We showed in the current work that it is only valid in the flavor
SU(3) symetric case, since the strange quark takes off a certain
amount of the down-quark contribution. 
\end{itemize}
In conclusion, our investigation emphasizes the importance of
considering sea quarks, especially strange quarks, in understanding
the gravitational form factors of the proton. It highlights the role
of the cosmological constant in the nucleon mass decomposition and its
implications for the mechanical properties of the proton. 

\vspace{0.5cm}

\textit{Acknowledgments} -- 
The work was supported by the Basic Science Research Program through
the National Research Foundation of Korea funded by the Korean
government (Ministry of Education, Science and Technology, MEST),
Grant-No. 2021R1A2C2093368 and 2018R1A5A1025563 (HYW and HChK).
This work was also supported by the U.S.~Department of Energy, Office
of Science, Office of Nuclear Physics under contract
DE-AC05-06OR23177~(JYK) and partially supported by the U.S. Department
of Energy, Office of Science, Office of Nuclear Physics under the
umbrella of the Quark-Gluon Tomography (QGT) Topical Collaboration
with Award DE-SC0023646~(JYK).    

\bibliography{SU3_GFFs}
\bibliographystyle{apsrev4-2}

\end{document}